\magnification 1192

%
\font\bigggfnt=cmr10 scaled \magstep 3
\font\biggfnt=cmr10 scaled \magstep 2
\font\bigfnt=cmr10 scaled \magstep 1

%
\leftskip .25 in
\rightskip .25 in
\vglue .85 in
\def\Par{\par\vskip 3 pt}
\def\ea{{\it et al.}}
\def\ie{{\it i.e.}}
\newbox\Ancha
\def\gros#1{{\setbox\Ancha=\hbox{$#1$}
   \kern-.025em\copy\Ancha\kern-\wd\Ancha
   \kern.05em\copy\Ancha\kern-\wd\Ancha
   \kern-.025em\raise.0433em\box\Ancha}}
\def\1i{\'{\i}}
\def\ni{\noindent}

\baselineskip 14.75 pt

\centerline{\bigggfnt Controlling a leaky tap }\Par

\vskip 30 pt

\centerline{\biggfnt Aquiles Ilarraza-Lomel\'{\i} }\Par
\vskip 6 pt
\centerline{\bigfnt Laboratorio de Sistemas Din\'amicos, Depto.\ de Ciencias B\'asicas}\Par
\centerline{\bigfnt Universidad Aut\'onoma Metropolitana-Azcapotzalco}\Par
\centerline{\bigfnt Apartado Postal 21-726, Coyoac\'an 04000 D.\ F., M\'exico}\Par
\vskip 15 pt
\centerline{\biggfnt C.\ M.\ Arizmendi  \hskip .1 cm and \hskip .14 cm A.\ L.\ Salas-Brito\footnote * 
{\hskip -4 pt\rm Corresponding author. On leave of absence from Laboratorio de Sistemas Din\'amicos, UAM-Azcapotzalco, e-mails: asb@hp9000a1.uam.mx or asb@data.net.mx. Postal address after April 6 1999: Apartado Postal 21-726 Coyoacan, M\'exico City  D.\ F.\ 04000 M\'exico} }\Par
\vskip 6pt
\centerline{\bigfnt Department of Physics, Emory University}\Par
\centerline{\bigfnt Atlanta, GA 30322, USA }\Par
\vskip 32 pt

\centerline {\bigfnt Abstract} \Par
\vskip 10 pt

We apply the Ott, Grebogy and Yorke mechanism for the control of chaos  to the 
analytical oscillator model of a leaky tap obtaining good results. We exhibit the 
robustness of the control against both dynamical noise and measurement noise.
A possible way of  controlling experimentally  a leaky tap  using magnetic-field-produced variations in the viscosity of a magnetorheological fluid is suggested. 
\vskip 22 pt

\vfill 
\noindent Classification numbers: 02.70+d, 47.20.Tg, 03.20.+i \Par 

\eject

The realization that the  majority of  natural phenomena are chaotic led to the  suggestion of 
chaotic behavior in the common household phenomena of a leaky tap or dripping faucet 
[1]---common perhaps, but not well understood, not even the process of  drop formation where the 
flow changes from a fluid mass to one or several falling drops; see, for example, [2--4] and the 
references therein. The first clear experimental evidence of such chaotic behavior was found by 
Shaw  and  his collaborators [5,6], further evidence was found a few years later by Wu and Schelly 
[7] and by N\'u\~nez-Y\'epez \ea\ [8]. Since then many experiments and theoretical works have 
established the system as a sort of paradigm for dissipative chaos [9--16].   \Par

 Shaw  proposed the first model for the process, a variable mass oscillator inspired 
in Rayleigh ideas [17, 18]. The model was actualized by S\'anchez-Ortiz 
and Salas-Brito (SOSB in what follows) [19, 20] and independently by D'Innocenzo and 
Renna [21--24] which, by changing the  breakup mechanism and the way of choosing  initial conditions, showed the broad range of behavior that can be achieved using the model and how it can be qualitatively related  to the experimental facts. A promising hydrodynamic model, aiming 
at a {\sl quantitative} agreement with the experiment and accounting for some of the topology 
transitions and the singularities in the phenomenom, has been recently put forward [25].  It must be clear that despite the enormous simplification in 
reducing a many-degrees-of-freedom fluid system to an one-dimensional model, there 
are many things that can be understood using the oscillator model since, basically due to 
dissipation,  the system restricts itself to  essentially one-dimensional attractors. Kiyono, Ishioka and Fuchikami have actually shown that the agreement between a model and the experimental results can be made quantitative by analysing the system from the perspective of fluid mechanics [25]. Using such ideas, Kiyono and Fuchikani have improved the relaxation oscillator model [26].  The oscillator idea
illustrates the important and sometimes underemphasized point  that for reproducing qualitative and even quantitative features of a chaotic system it is usually not necessary to use very complex models. \Par

Moreover, the leaky tap and the oscillator model have  been used as a sort of role 
model to simulate other complex phenomena [27]; furthermore, given the similarity between certain 
of their features [28], they can be of help in modelling the comparable-to-chaotic heartbeat
behavior [28--30].  The experimental control of a leaky tap can then be of importance as a 
testing ground for certain ideas. For instance, a cardiac
MR imaging technique has been proposed which employs time series forecast
and standard methods for the analysis of chaos on heartbeat time
series. Using such concepts, new pacemakers are being investigated which use control
techniques to correct arrhythmic behavior of the heart while minimizing
their intervention and battery consumption. See [31] and the references therein. \Par 

Our aim in this work is to apply a model independent chaos control technique to 
the SOSB equations in the analytical approach of D'Innocenzo and Renna [24]; 
 we then  suggest an experimentally-realizable scheme for 
the control of an actual dripping faucet. We carry out the control using the 
Ott, Grebogy and Yorke strategy  (OGY in what follows) [32]; the  advantages of 
the OGY method is that  it does not need a detailed knowledge or model of the 
phenomena and  it uses the chaotic behavior itself as the mechanism of control. 
We have found that it is possible to stabilize the SOSB model around one of its 
unstable equilibrium points and that such control is robust (between certain 
limits) against external perturbations; this is obviously a good feature with  
an  experiment in mind. The control is accomplished by adjusting the parameter of 
the SOSB model analogous to the viscosity of the leaking fluid. \Par

Let us begin by reviewing the  SOSB relaxation oscillator model [33,34]. 
The starting equations, in nondimensional coordinates, are [19,20]

 $$   \eqalign {
    {dx \over dt} &= y, \cr
    {dy \over dt} &= {-{1\over m}(x+ \beta \, y) } + g, \cr
    {dm \over dt} &= f,
}   \eqno(1) \qquad
$$

\noindent where, $\beta$, $g$ and $ f$ are parameters modelling viscosity, external force (gravity) 
and water inflow, respectively, and $m$ is the mass. \Par

 We here, following the analytic approach of D'Innocenzo and 
Renna [23] and instead of studying directly numerical solutions to the set of 
equations (1), use a sort of approximate solution to it, namely

$$ x(t)= \left[A\sin \omega(t)\, t + B \cos \omega(t)\, t\right]\exp(-\gamma(t)\, 
t)+{m(t)\,g}, \eqno(2) \qquad $$

\noindent where $m(t)=m_0 +f (t-t_0)$, $\gamma(t)\equiv \beta/m(t)$ and $\omega^2(t)
\equiv 1/m(t)$; equation (2)  together with the following proviso for the drop breakup: when the position of the oscillator reaches the meniscus length (normalized such that $x_c=1$)  a drop is forced to detach, provoking an abrupt diminution
in the oscillator mass by the quantity [19,20]

$$ \Delta m= h\,\, m(t_c)\, y(t_c), 
\eqno (3)\qquad $$ 

\noindent $h$ being a model parameter and where we have set $t\equiv t_c$.  So before using (2) 
again, we have to reset the starting value of the oscillator mass to the new value 
$m_0=m(t_c)-\Delta m$;  this scheme substitutes for the missing singularity-forming description of 
drop detachment [19--21]. The analytic model (2) was first proposed by D'Innocenzo and Renna [21] and has been recently used to reproduce [24] the closed loop attractors and the Hopf bifurcations experimentally observed in a leaky tap [7, 15--16].\Par

 We further assume that the clock resets every 
time a drop breaks off, \ie\ we take $t_0=0$ for every new drop. To give a 
criterion for the initial position of the next drop we use a sort of amorphous
 drop model, loosely inspired in Eggers study [2]. This has lead us to propose a way of choosing the initial conditions after breakup  such that 

$$x_0=\exp(-\Delta m),  \eqno(4) \qquad$$

\noindent this amorphous drop mechanism guarantees that  $0<x_0<1$. Furthermore, the initial velocity of 
the remaining fluid can always be taken as $y_0=y(t_c)$ simply considering that the 
mass of the remaining fluid, {\sl plus} the effect of the mass inflow,  produce an effective mass 
term very large compared to $\Delta m$ on which the snapping back of the system has a negligible 
effect. The use of the amorphous model (4) had, in fact,  its origin in our 
attempts at  using the spherical drop model of D'Innocenzo and Renna [21--23] that, sometimes, 
resulted in detached drops much larger than the normalized meniscus lenght.\Par

 Our analytical SOSB model is hence
specified  by equations (2) and (3) together with the amorphous drop scheme for the initial 
position of the next drop. Such mechanism of releasing drops, besides giving 
reasonable values for the initial positions generates important correlations between
succesive drops. \Par

 Although it is known that the results obtained from the model vary 
greatly according to the specific mechanism employed for simulating the drop detachment [23], we 
here limit ourselves to quote results from the amorphous drop mechanism (4). We point out that we have tried 
different mechanisms and other variations of the SOSB model getting the same general conclusion 
[33]. \Par

 Using a  modified Newton-Raphson method [35] for getting $t_c$ from the dripping condition
$x(t_c)=1$, a very efficient method of simulating the dripping  tap behavior is obtained 
[21,22]. We should point out that the analytical SOSB equations (2)
 and (3) are appropriate for modelling leaking from relatively big (with diameters larger
than $\sim1/4$ cm) faucets, since for such diameters  the drop dynamics is mainly governed by the 
center of mass motion of the hanging fluid  [2, 17].  We are then
capable of accumulating a large number of drip intervals $t_n\equiv t^{(n)}_c$ for analysis. The
drip intervals, \ie\ the time spans separating a drop from  the next one,  have become the 
standard  variables used in all leaky tap studies.\Par

 Bifurcation diagrams---dripping spectra in the 
terminology of Wu and Schelly [7]---, time series and return maps $t_{n+1}$
versus $t_n$, illustrating the results that can be obtained from the analytical SOSB
equations, are shown in Figure 1 and 2 [33, 34]. Figure 1 shows both a bifurcation diagram (figure 1a) and a time series (figure 1b) taken from the zone we want to control.
Notice also that at the parameter values  
($g=0.4$, $h=0.3$ and $\beta=3.01$) used the system undergoes a period-doubling sequence, shows evidence
 of crisis (figure 1a) and behaves intermittency (figure 1b) [18]. The horizontal dashed line in figure 1b simply marks the 
drip-interval  at the unstable period-1 point ($t_c=t_F=0.2858$) we aim to control (but it does {\bf not} necessarily coincide with the seemingly intermittent state appearing in figure 1b). The bifurcation 
diagram also shows that, in the conditions of figure 1a,  dripping is interrupted by 
``continuous flow'' at $f > 8.5002$, that 
is, at greater $f$-values the oscillator position always remains larger than the meniscus lenght 
after the detachment of the first drop [19,20]. \Par

Let us mention that the parameter values used in our dicussion do not attempt to be typical of an experimental situation, rather they were chosen  with the sole purpose of illustrating the OGY mechanism as applied to the system. We must mention though that the results were checked for other values of the parameters, that is, for chaotic attractors of different sorts, always  obtaining similarly good results---although the dynamics may be different but still chaotic. The method even allowed us to control the system in an unstable  period-10  cycle [33].\Par

 Figure 2 shows the reconstructed attractor that exists at the unstable fixed point location. 
The fixed point is shown as a small black dot in the figure 2 inset. Notice that the 
reconstructed attractor has a complex  structure [20]  which can be regarded as  
difficulting the precise location, and hence the control, of the unstable fixed point. 
Nevertheless, the location of such fixed point and its control are easily achieved and are basically
limited  by the precision of our computations.   \Par

 To identify the unstable fixed point orbit (shown in Figure 2) in the otherwise chaotic 
attractor-dominated dynamics, we simply acknowledge that every chaotic system has an agglomeration 
of unstable periodic orbits embedded almost anywhere. Hence, 
there must be a fixed point whenever an attractor crosses the line of the identity
($t_{n+1}=t_n$) in a return map.  Using the  time series, we can numerically identify the 
unstable fixed point at $t_c=t_F=0.2858$  present in the attractor shown in figure 2.  \Par

Around the unstable fixed point ${\bf X}_F=(t_n=t_F,\, t_{n+1}=t_F)$ in the return map, with the help of the time series, we can use a locally linear dynamics [32] to describe the system

$$  D\cdot ({\bf X}_n-{\bf X}_F)= {\bf X}_{n+1}-{\bf X}_F,    \eqno (5) \qquad $$
 
\noindent where $D$ is a $2\times2$ matrix and ${\bf X}_n$ is the vector with components
$(t_{n+1}, t_n)$. With the local dynamics (5), it is then a simple matter to calculate the
normalized $D$-eigenvectors, ${\bf e}_s$, ${\bf e}_u$, and its corresponding eigenvalues,   
$\lambda_s$, $\lambda_u$, associated with its stable and  unstable manifolds. From 
them, we can evaluate also the contravariant vector  associated with the unstable manifold as 
the vector ${\bf f}_u$, for which $ {\bf f}_u\cdot{\bf e}_u=1$ and ${\bf f}_u\cdot{\bf e}_s=0$ 
holds [32]. \Par

To control the system we have chosen to adjust the viscosity parameter $\beta$; we choose $\beta$ and not the seemingly more natural fluid inflow, because we have in mind a magnetorheological fluid in which the viscosity can be varied using an easily tuned magnetic field and more important because  it is  rather difficult to control $f$ with confidence, in our conditions at least. We thence need to evaluate ---numerically from the time series--- the sensitivity of the model to changes in the viscosity parameter $\beta$ respect to a fiducial value $\beta_0$, as

$$ {\bf s}=  \left.{\partial {\bf X}_F \over \partial \beta}\right|_{\beta_0}. \eqno (6) \qquad$$

With the above information we let the system run and apply the control every time the drip interval is within an appropriate fixed-point neighborhood; such neighborhood is specified through the inequality

$$\left|\left({\bf X}_n-{\bf X_F}\right)\cdot {\bf f}_u\right| < \xi_*,  \eqno(7)\qquad$$

\noindent where

$$ \xi_* = \left|\delta \beta_*\, ({\bf s} {\gros\cdot} {\bf f}_u)\, 
\left(1-{1\over \lambda_u}\right)\right|, \eqno (8) \qquad $$

\noindent and $\delta \beta_*$ is the maximum value allowed (see below for the value set) for 
changes in the viscosity para\-meter---we pinpoint that (8) is only valid as a first-order approximation.
Once with the control working and after a brief transitory the system never gets far from 
${\bf X_F}$, as figures 3a and 3b show.
 Notice that, again, we choose controlling the dynamics around the interval
defined by (7) and (8); this is quite appropriate with an experimental situation in mind. 
The scheme described is simply the OGY control method, forcing the system to evolve towards the 
unstable direction by changing slightly the value of the $\beta$ parameter [32]. \Par

The OGY scheme, applied to the model dynamics lead to the results  shown in Figure 3. 
The results correspond to an unstable fixed point $t_F=0.2858$, found within the chaotic
attractor at the parameter values $f=8.49$, $\beta=3.01$, $g=0.4$ and 
$h=0.3$; the behavior is intermittent at these parameter values (figure 1b). The control 
parameters used are  $\delta \beta_*=1.3$, $\xi_*=0.008$, 
${\bf s}= (-0.0021, -0.0021)$, and the 
unstable  eigenvalue is $\lambda_u=-1.24$; the unstable manifold is 
associated with the contravariant vector

$$ {\bf f}_u=\pmatrix{-0.8940\cr 0.7020}. \eqno(9) \qquad$$

\noindent The explicit expressions given above for the unstable and stable eigenvectors guarantee 
that we are not in an homoclinic tangency point of the attractor [36] which is an unsuitable point for applying the OGY method. In Figure 3a we can observe a consistent estabilization of the system after the application of 
the control in $n=1000$; it takes less than 150 drippings to get the system into the fixed point. The control is released after drop 3000 and chaos sets 
in inmediately; it is again applied at $n=5000$, and 150 or so drippings ahead 
the system becomes periodic again. Further information about the approach to the fixed point 
once the control is applied, can be obtained from a return map of the process; this is shown
in Figure 3b. The spiralling approach, clearly shown in the inset on figure 3b, to the 
fixed point seems to be typical. We have to conclude then that with no perturbations present the control seems to work very well. \Par

An appropriate question is what happens if there are extra random perturbations. Such perturbations are 
expected to occur in any experimental realization of the leaky tap. In what follows we first analyse 
the effect of  random noise
superimposed to the value of the parameter $f$. We should term this the case of dynamical noise, since experimentally it arises from the impossibility of keeping perfectly fixed the inflow. In fact, we choose such parameter to illustrate the robustness of the control precisely because the fluid flow 
into the tap is a difficult variable to keep fixed in an experimental situation 
[8, 12]. This also explains why we do not considered  proposing an experimental mechanism of control using the inflow $f$---though the control is equally easy to achieve adjusting $f$ but {\sl in the model} [33]. The random perturbation is applied as 
$f=f_0 + \delta f$ in the model equations, where $f_0$ is now the fiducial value (\ie\ $f_0=
8.49$, as in figures 2 and 3) and $\delta f$ is a uniformly distributed random variable in [$-0.0043, 0.0043$]. 
We have to be sure that such random perturbation does not significantly change the dynamics since the $f$-width of the chaotic zone is small. In 
figure 4 we show, as an example, a time series and a return map with fiducial values of the parameters  as in Figure 2, but with the 
random perturbations applied to $f$ allowing for variations  up to $1 \%$ of its fiducial value 
(note that  such variations represent $10\%$ of the total width of the chaotic zone). It can be seen that the  
dynamics just  become fuzzier compared to the original unperturbed case.  Notice also that the 
system does not permit imposing larger variations in $f$, otherwise we will leave the rather 
small chaotic zone (the $f$-width of that zone is 0.035, as shown in Figure 1) and the dynamics would then be 
drastically changed.\Par

What happens with the control scheme turned on? The control was applied without modification to the 
perturbed system and the results show that the scheme is rather robust under  random perturbations in  $f$. 
Figure 5 show time series and return maps of the randomly perturbed system under control. Notice that the 
spiral approach to the fixed point has become an ellipsoidal blob of points; this figure roughly corresponds 
to the area of the interval (8) in the reconstruction space. Incidentally, notice that figures 4b and  5b 
also illustrates the predicted noise induced attractor deformation and elongation [36]  recently observed in 
periodically driven non-linear electric circuits [37]. Figure 5b also illustrates that the control 
effectively stabilizes the system to a neighborhood of the fixed point, not allowing vagaries larger than the maximum size of the control zone.\Par

But the lack of control in $f$ is not the only perturbation worth of analysing. The unavoidable 
uncertainties 
in the time measurements, that is, what we can term the case of measurement noise, and the problem of lost drops are also important.  We simulate such behavior by randomly perturbing the values of the drip intervals calculated from the model. We consider thus drip 
intervals 
$t_n= t_n^{(0)}+\delta t_n$ where $t_n^{(0)}$ is the drip interval calculated from (2) and (3) and 
$\delta t_n$ is a random variable with, again, a uniform distribution. What we found using these 
`measured' 
drip intervals is that, if the uncertainty introduced by the random noise is larger than $\sim 
0.5 \%$ of the maximum value of the drip interval, despite the intended control, the system 
exhibits little 
but noticeable chaotic bursts. The bursts become larger as the uncertainty   
grows until the control is completely lost. This is shown for succesively larger values of the perturbation 
in Figures 6a, 6b and 6c. In the last of the time series shown (Fig.\ 6c, with a  perturbation of $\simeq 10\%$ of 
the total $t$-width ($\simeq 0.02995$) of the chaotic zone) traces of the control still are noticeable but overall the system is 
destabilized and chaotic. Such behavior can be easily understood when it is considered  that at such 
uncertainties it is no longer possible to tell apart a drop from the adjacent ones. In this measurement noise case then, it is possible to quote the noise values  which the control mechanism found acceptable, whereas in the previous dynamical noise  case it was not possible due to the small $f$-width of the chaotic zone [7, 12, 15]; in the dynamical noise case the system would no longer be within the chaotic zone before the control collapses by increasing the  noise level. That dynamical noise could throw the system out from the chaotic regime, can also happen in the experiment [8] but, in  such a case, the large fluctuations in $f$ would simply mean that the experiment is not working properly.  \Par

 In all the examples given,  the control procedure used is applied using 
the approximate linear dynamics calculated from the unperturbed system, which is a sort of idealistic case. In a more realistic situation, the local dynamics will be evaluated from the actual measurements and this would improve the control. \Par

The results of our computations with the relaxation oscillator SOSB equations hint towards a 
control technique  applicable to the leaky tap in an experimental situation. 
We require a system with at least a parameter allowing  quick adjustment and  quick response times as 
compared with typical drip intervals; typical values of $t_n$ in an experiment are 
of the order of 100 ms [8, 12, 16]. The chosen control variable 
should allow faster response that this typical value. We have thought therefore on adjusting 
the fluid viscosity because the inflow $f$ is not easy to control, at least from the viewpoint of our Laboratory. On the other hand, common fluids (water is the working fluid in every experiment 
performed to date) are very difficult to change their viscosity 
excepting with changes in temperature, but this is not easy to accomplish in the required 
circunstances. Had we thought of changing the temperature of the water, we would need rather 
large changes which would also change other system parameters---as the diameter
 of the nozzle---and temperature would not be so easy to control. \Par

 To overcome such  anticipated difficulties, we propose the use, instead of the customary water,  of an oil-based magnetorheological fluid as the leaking fluids in the system. Such  fluids
are easy to obtain, have response times of 2 or 3  milliseconds
 ---almost an order of magnitude below the typical drip intervals in water and the drip
 intervals are larger in the magnetorheological fluid given its greater viscosity. Besides, they can quickly
change their viscosity for up to a $10^6$ factor [35] (though for our purposes we do not need such huge changes) simply applying a magnetic field, which is also rather easy to adjust. Giving such characteristics, we think that the method would allow an excellent control.\Par

In summary, we have applied succesfully the OGY control method to the SOSB leaky tap model, investigating the possible ill-effects of random noise on the water inflow into the tap and on the drip intervals. We have found the the control procedure is effective up to noise to signal ratios of the order of $\sim 10\%$. We should also mention that all the computations reported in this article were carried out in fortran 77 using a PC workstation  running under Linux. \Par

To finalize, we have to say that the study in [25] has been further used to improve the oscillator model. The main change has been the use of a mass-dependent elastic `constant' $k$ for the spring [26] (which we here normalized to 1); with the proper indentification of the model parameters a  very good agreement  with the experimental values [6--10] is found. This  adds to the usefulness of the oscillator model as it is further illustrated by this contribution.\Par

\noindent {\bf Acknowledgements.}\Par

\noindent This work has been partially supported by CONACyT (grant 1343P--E9607) and by PAPIIT-UNAM (grant IN--122498) which partially financed ALSB's trip to Atlanta. ALSB and CMA want to thank Emory University and particularly Professor F.\ Family for all the support and the warm hospitality in Atlanta. We also acknowledge helpful discussions with H.\ N.\ N\'u\~nez-Y\'epez,  M.\ N.\ Popescu, G.\ I.\ S\'anchez-Ortiz and E.\ Guillaum\1in-Espa\~na, the help of J.\ Estrada-D\1iaz, Jorge Reyes-Iturbide and the cheerful mood of all the team at LSD.  G.\ Hentschel suggested that calling a plumber is perhaps the best way for controlling a leaky tap and we had to agree with him.
This work owes a great deal  to the encouragement of Q.\ Tavi, K.\ Hryoltiy,  M.\ Chiornaya, F.\ C.\ Minina, L.\ Bidsi, G. Abdul II, U.\ Kim and C.\ Srida. Last but not least, we want to dedicate the article to the memory of our beloved friend L.\ Tuga. \Par

\vfill
\eject

 \noindent {\bf References}. \Par

\noindent [1] O.\ E.\ R\"ossler, {\it Synergetics, A workshop}, H.\ Haken, ed.\ Springer Berlin p.\ 174 (1977).\Par 

\noindent [2] J.\ Eggers, lanl archive preprint ``Singularities in droplet pinching with vanishing viscosity'' 
(chao-dyn/9705005, v2, 1999).\Par

\noindent [3] X.\ D.\ Shi, M.\ P.\ Brenner, S.\ R.\ Nagel, Science {\bf 265} (1994) 219.\Par

\noindent [4] R.\ E.\ Goldstein, R.\ I.\ Pesci and M.\ J.\ Schelly, Phys.\ Rev.\ Lett.\ {\bf 70} (1993) 3043.\Par 

\noindent [5] R.\ Shaw, {\it The dripping faucet as a model chaotic system} (Aerial Press, Santa Cruz USA, 1984). \Par

\noindent  [6] P.\ Martien, S.\ C.\ Pope, P.\ L.\ Scott and R.\ S.\ Shaw,
Phys.\ Lett.\ {\bf 110A} (1985) 399. \Par

\noindent [7] X.\ Wu and Z.\ A.\ Schelly, Physica D {\bf 40} (1989) 433.
\Par

\noindent  [8] H.\ N.\ N\'u\~nez-Y\'epez, A.\ L.\ Salas-Brito,
C.\ A.\ Vargas,  and L.\ Vicente, Eur.\ J.\ Phys.\ {\bf 10} (1989) 99; reprinted in L.\ Lam, {\it Nonlinear physics for beginners} (World Scientific, 1998). p.\ 104.\Par 

\noindent  [9] R.\ F.\ Cahalan, H.\ Leidecker and G.\ D.\ Cahalan, Comp.\
Phys.\ {\bf 3} (1990) 368. \Par

\noindent [10] J.\ Austin, Phys.\ Lett.\ A {\bf 155} (1991) 148. \Par

\noindent  [11] G.\ I.\ S\'anchez-Ortiz, Tesis de Licenciatura, {\it El grifo
goteante:  estudio num\'erico de un modelo mec\'anico} (FCUNAM, M\'exico
D.\ F., 1991).  \Par

\noindent  [12] H.\ N.\ N\'u\~nez-Y\'epez, C.\ Carbajal, A.\ L.\ Salas-Brito,
C.\ A.\ Vargas and L.\ Vicente,  in {\it Nonlinear phenomena in fluids,
solids and other complex systems}, P.\ Cordero and B.\ Nachtergaele eds,
(Elsevier, Amsterdam, 1991) p.\ 467. \Par

\noindent  [13] P.\ M.\ C.\ de Oliveira and T.\ J.\ P.\ Penna, J.\ Stat.\ Phys.\ {\bf 73} (1993) 789.\Par

\noindent [14] P.\ M.\ C.\ de Oliveira and T.\ J.\ P.\ Penna, Int.\ J.\ Mod.\ Phys.\ C {\bf 5}  (1994) 997.\Par

\noindent  [15] J.\ C.\ Sartorelli, W.\ M.\ Gon\c{c}alves and R.\ D.\ Pinto,
Phys.\ Rev.\ E {\bf 5} (1994) 3963. \Par

\noindent [16] R.\ D.\ Pinto, W.\ M.\ Gon\c{c}alves, J.\ C.\ Sartorelli, and M.\ J.\ de Oliveira, Phys.\ Rev.\ E {\bf 52} (1995) 6896.\Par

\noindent [17] J.\ W.\ S.\ Rayleigh, Proc.\ London Math.\ Soc.\ {\bf 4} (1878) 10.\Par

\noindent [18] J.\ W.\ S.\ Rayleigh, {\it The theory of sound} (Dover, New York, 1945) \S 364.\Par

\noindent  [19] G.\ I.\ S\'anchez-Ortiz and A.\ L.\ Salas-Brito,  Phys.\ Lett.\ A {\bf 203} (1995)  
300. \Par 

\noindent  [20] G.\ I.\ S\'anchez-Ortiz and A.\ L.\ Salas-Brito, Physica D  {\bf 89} (1995)  151. \Par

\noindent [21] A.\ D'Innocenzo and L.\ Renna, Phys.\ Lett.\ A {\bf 220} (1996a) 75.\Par

\noindent [22] A.\ D'Innocenzo and L.\ Renna, Int.\ J.\ Theor.\ Phys.\ {\bf 35} (1996b) 941.\Par

\noindent [23] A.\ D'Innocenzo and L.\ Renna, Phys.\ Rev.\ E {\bf 55} (1997) 6776.\Par

\noindent [24] A.\ D'Innocenzo and L.\ Renna, Phys.\ Rev.\ E {\bf 58} (1998) 6847.\Par

\noindent [25] N.\ Fuchikami, S.\ Ishioka and K.\ Kiyono, lanl archive preprint ``Simulations of a dripping faucet'' (chao-dyn/9811020, 1998).\Par

\noindent [26] K.\ Kiyono and N.\ Fuchikami, lanl archive preprint ``Dripping faucet dynamics clarified by 
an improved mass-spring model'' 
 (chao-dyn/9904012, 1999).\Par

\noindent [27] D.\ N.\ Baker, A.\ J.\ Klimas, R.\ L.\ McPherron and J.\
Buchner, Geophys.\ Res.\ Lett.\ {\bf 17} (1990) 41. \Par

\noindent [28] T.\ J.\ P.\ Penna, P.\ M.\ C.\ de Oliveira, J.\ C.\ Sartorelli, W.\ M.\ Gon\c calves and R. D. Pinto, Phys.\ Rev.\ E {\bf 52} (1995) R2168. \Par

\noindent [29] C.-K.\ Peng, J.\ Mietus, J.\ M.\ Haussdorff, S.\ Havlin, H.\ E.\ Stanley, and A.\ L.\ Goldberger, Phys.\ Rev.\ Lett.\ {\bf 70} (1993) 1343.\Par

\noindent [30] K.\ Otsuka, G.\ Cornelissen, F.\ Halberg, Clin.\ Cardiol.\ {\bf 20} (1997) 631; J.\ Kanters, N.\ Henrik, H.\ Pathlou and E.\ Agner, J.\ Cardiovascular Electrophysiology {\bf 5} (1994) 591; A.\ Garfinkel, M.\ L.\ Spano, W.\ L.\ Ditto and J.\ N.\ Weiss, Science {\bf 257} (1992) 1230.\Par

\noindent [31] G.\ I.\ S\'anchez-Ortiz, D. Rueckert y P. Burger, Medical Image Analysis {\bf 3} (1999) 77.\Par

\noindent [32] E.\ Ott, C.\ Grebogi, and J.\ A.\ Yorke, Phys.\ Rev.\ Lett.\ {\bf 64}, 1196 (1990).\Par
\Par

\noindent [33] J.\ Estrada-D\1iaz, Proyecto Terminal, {\it An\'alisis din\'amico-param\'etrico del sistema goteante}, (UAM-Azcapotzalco, M\'exico City, 1998); A.\ Ilarraza-Lomel\1i, Pro\-yecto Terminal, {\it Control del grifo goteante usando el fen\'omeno magnetorreol\'ogico}, (UAM-Azcapotzalco, M\'exico City, 1999).\Par 

\noindent [34] A.\ Ilarraza-Lomel\1i,  J.\ Estrada-D\1iaz, C.\ M.\ Arizmendi, A.\ L.\ Salas-Brito to be submitted (1999).\Par

\noindent [35] W.\ H.\ Press, S.\ A.\ Teukolsky, W.\ A.\ Vetterling, and B.\ P.\ Flannery, {\it Numerical Recipes in Fortran} (Cambridge University Press, Cambridge, 1992), ch.\ 9.\Par

\noindent [36]  L.\ Jaeger and H.\ Kantz, Physica D {\bf 105} (1997) 79.\Par

\noindent [37] M.\ Diestelhorst, R.\ Hegger, L.\ Jaeger, H.\ Kantz and R.-P.\ Kpasch, Phys.\ Rev.\ Lett.\ {\bf 82} (1999) 2274.\Par

\noindent [38] R.\ E.\ Rosensweig, {\it Ferrohydrodynamics} (Dover Publications, New York, 1998).\Par

\vfill
\eject

\ni{\biggfnt Figure Captions.}\Par

\noindent Figure 1.\par
Illustration of the behavior predicted by the relaxation oscillator model for the leaky tap.  Many  other examples of the possible behavior can be found in [21--25, 33,34].\par

\noindent 1a.   Bifurcation diagram at $\beta=3.01$, $g=0.4$, $h=0.3$ as $f$ is varied. Notice that beyond $f=8.500019$ the dripping stops and ``continuous flow'' sets in. The vertical dashed line passing through $f=8.49$ marks the zone to control. Notice that the chaotic zone after the period doubling bifurcations extends roughly from 8.465 to 8.500, a total $f$-width of 0.035. \par

\noindent 1b. Time series in the zone we want to control ($\beta=3.01$, $g=0.4$, $h=0.3$ and $f=8.49$). Notice the signals of intermittency. The thin dashed line $t=0.2858$ corresponds to the unstable fixed point we intend to stabilize. Let us emphasize that we do not intend to control the  intermitent orbit which seems to coincide with the selected unstable fixed point.\Par

\noindent Figure 2.\par

\noindent Return map $t_{n+1}$ {\sl vs.\ } $t_n$ showing the unstable fixed point at $t_F=0.2858$. The inset is a blow up of the square  neighborhood depicted around the fixed point. Such unstable orbit is pointed to by a black arrows and marked by a black dot in the inset. As you can notice, the attractor has a complex structure composed of at least two very close sheets. The fixed point lays in the innermost sheet of the reconstructed attractor. \Par

\noindent Figure 3.\par

\noindent Effect of the OGY scheme on the dynamics; compare with figures 1b and 2.\par

\noindent 3a.  Time series of drip intervals for the unperturbed model with the control turned on and off. At $n=1000$ the control is applied, it takes roughly 150 drops for the system to be stabilized into the unstable fixed point at $t=t_F=0.2858$. At $n=3000$ the control is released and chaotic behavior sets in immediately. At $t=5000$ the control is applied again. \par
  
\noindent 3b. Return map of the control process. Notice the spiral approach to the unstable fixed point when the control is turned on. The inset is a blow up of the region, exactly the same as described in figure 2, around the unstable fixed point.\Par

\noindent Figure 4.\par

\noindent The SOSB model in the  presence of random perturbations applied to the value of $f$. The noise level is $\simeq 10 \%$ of the $f$-width of the chaotic zone. In this case, it is not possible to increase the noise for testing the robustness of the control without first leaving the rather small ($f$-width $=$ 0.035) chaotic zone.\par

\noindent 4a. Time series of the drip intervals with  random noise superposed on $f$. A comparison with figure 1b may show that the dynamics gets fuzzier. \par

\noindent 4b. Return map of the zone to be controlled with random noise on $f$ superposed. Compare to figure 2. Notice the deformation and the elongation of some parts of the reconstructed attractor induced by the applied random noise [36,37]. The fuzziness mentioned in 4a becomes evident.\Par

\noindent Figure 5.\par

The OGY scheme applied to the SOSB model in presence of random noise on $f$. The noise level is $\simeq 10 \%$ of the $f$-width of the chaotic zone.\Par

\noindent 5a Time series of the $f$-perturbed SOSB leaky tap model, with the control turned on at $n=5000$. Despite the noise the system stabilizes around the unstable fixed point.

\noindent 5b. Return map
of the system with the control turned on. The espiral approach to the fixed point becomes an approximately elliptical region where the system gets controlled. \Par

\noindent Figure 6.\par

Effect of random noise applied to the drip intervals. Notice that in the conditions of figure 6b it begins to be difficult to tell apart a drop from adjacent ones and that, in the conditions of figure 6c, it is almost not possible. \par

\ni 6a. The noise level is here $0.5 \%$ of the maximun range allowed for $t$. Control is still rather good.\par

\ni 6b. The noise level is $0.75 \%$ of the maximun range in $t$. The bursts of chaos were control is lost are  evident, control is also present though far from perfect.\par

\ni 6d.  The noise level is $1 \%$ of the maximum range in $t$.
Traces of  control  still remain but it is almost completely lost.

\end